\documentclass[twocolumn,prl,aps,superbib,tightenlines,floatfix]{revtex4}
\usepackage{amsfonts,amsmath,amssymb,amsthm,bm}
\usepackage{graphicx}
\begin{document}
\title{Effect of electron-phonon scattering on shot noise in nanoscale
junctions}
\author{Yu-Chang Chen}
\author{Massimiliano Di Ventra}
\email[]{diventra@physics.ucsd.edu}
\affiliation{Department of Physics, University of California, San
Diego, La Jolly, CA 92093-0319}
\begin{abstract}
We investigate the effect of electron-phonon inelastic scattering
on shot noise in nanoscale junctions in the regime of
quasi-ballistic transport. We predict that when the local
temperature of the junction is larger than its lowest vibrational
mode energy $eV_c$, the inelastic contribution to shot noise
(conductance) increases (decreases) with bias as $V$ ($\sqrt{V}$).
The corresponding Fano factor thus increases as $\sqrt{V}$. We
also show that the inelastic contribution to the Fano factor
saturates with increasing thermal current exchanged between the
junction and the bulk electrodes to a value which, for $V>>V_c$,
is independent of bias. A measurement of shot noise may thus
provide information about the local temperature and heat
dissipation in nanoscale conductors.
\end{abstract}
 \maketitle It is an
established fact that for systems with dimensions much longer than
the inelastic mean free path $\lambda _{ph}$ (e.g. a macroscopic
sample) steady-state zero temperature current fluctuations (shot
noise) are suppressed by electron-phonon
scattering~\cite{buttiker,shimizu,nagav}.
Similarly, for metallic diffusive wires with length much smaller than $%
\lambda _{ph}$ (and smaller than the electron-electron scattering
length), the Fano factor (i.e. the ratio between shot noise and
its Poisson value, $2eI$, where $e$ is the electron charge and
$I$ is the current of the
system) equals $1/3$ and is not affected by inelastic processes~\cite%
{beenakker}. Systems of nanoscale dimensions may not fall in either
one of the above cases. In this instance each electron, on average,
releases only a small fraction of its energy to the underlying
atomic structure during the
time it spends in the junction, making transport quasi-ballistic~\cite%
{todorov1,todorov2,todorov3,chen23,troisi,chen05,yang}. However, the current density and,
consequently, the power per atom are much larger in the junction
compared to the bulk. This leads to heating and inelastic features
in the differential conduction which are indeed observed in
experiments with metallic quantum point
contacts~\cite{agrait,mizo,halbritter,smitnanotech} and molecular structures~\cite%
{Kushmerick,Wang2,natel,chen23,chen05} as a direct consequence of the interplay
between electron and phonon statistics~\cite{Heidi}. For these systems it is
therefore not obvious what is the effect of inelastic scattering
on shot noise.

In this Letter we show analytically that shot noise in
quasi-ballistic nanoscale junctions is enhanced by inelastic
scattering whenever electrons have enough energy to excite the
phonon modes of the junction. The current instead decreases. As a
consequence, the Fano factor increases. We find it increases with
bias as $\sqrt{V}$ when the local temperature of the junction is
larger than its lowest vibrational mode energy $eV_c$. We also show that
with increasing thermal current carried away from the junction to
the bulk electrodes, the inelastic contribution to the Fano factor
converges to a minimum value independent of bias for $V>>V_c$. A measurement of the Fano factor may thus provide
information about the local temperature and heat dissipation in
nanoscale conductors. Transport in a model atomic gold point
contact will be used to illustrate these findings.

Since the dimensions of the junction are
much smaller than $\lambda _{ph}$ (and the observed inelastic
features in quasi-ballistic systems are very
small~\cite{agrait,Kushmerick,Wang2}) first-order perturbation
theory in the electron-phonon coupling captures the dominant
contribution to inelastic scattering. This is the contribution we
calculate in this paper.

Let us assume that the junction is connected to two biased bulk
electrodes. The electronic states of the full system are thus
described by the field operator $\hat{\Psi}=\sum_{E,\alpha
=L,R}a_{E}^{\alpha }\Psi _{E}^{\alpha }\left(
\mathbf{r,\mathbf{K}_{\Vert }}\right) $, constructed from the
single-particle wave functions $\Psi _{E}^{L(R)}\left( \mathbf{r,\mathbf{K}%
_{\Vert }}\right) $ and annihilation operators $a_{E}^{L(R)}$
corresponding
to electrons propagating from the left (right) electrode at energy $E$. $%
\mathbf{K}_{\parallel }$ is the transverse component of the momentum~\cite%
{diventra}. We also assume that the electrons rapidly thermalize
into the bulk electrodes so that their statistics are given by the
equilibrium Fermi-Dirac distribution, $f_{E}^{L(R)}=1/(\exp
[(E-E_{FL(R)})/k_{B}T_{e}]+1) $ in the left (right) electrodes
with local chemical potential $E_{FL(R)}$, where $T_{e}$ is the
electronic temperature. In the following we will assume that $T_{e}=0$ K~\cite{thermal}, and the
left electrode is positively biased so that $E_{FL}<E_{FR}$. The
stationary scattering states $\Psi _{E}^{L(R)}\left(
\mathbf{r,\mathbf{K}_{\Vert }}\right) $ are eigenstates of an
effective single-particle Hamiltonian $H_{e}$ which may be
computed, e.g., using a scattering approach within the static
density-functional theory of many-electron
systems~\cite{diventra}. The combined dynamics of electrons and
phonons is described by the Hamiltonian (atomic units will be used
throughout this paper)~\cite{chen23}

\begin{equation}
H=H_{e}+H_{ph}+H_{e-ph},  \label{Ham}
\end{equation}%
where $H_{ph}=\frac{1}{2}\sum\limits_{i,\mu \in vib}\dot{q}_{i\mu }^{2}+%
\frac{1}{2}\sum\limits_{i,\mu \in vib}\omega _{i\mu }^{2}q_{i\mu
}^{2}$ is the phonon contribution, with $q_{i\mu }$ the normal
coordinate and $\omega _{i\mu }$ the normal frequency of the
vibration labeled by the $\mu $-th component of the $i$-th ion.
$H_{e-ph}$ describes the electron-phonon interaction and has the
following form~\cite{chen23}

\begin{align}
H_{e-ph}& =\sum_{\alpha ,\beta }\sum_{E_{1},E_{2}}\sum_{i\mu ,j\nu
\in vib}
\notag \\
& \sqrt{\frac{1}{2\omega _{j\nu }}}A_{i\mu ,j\nu
}J_{E_{1},E_{2}}^{i\mu ,\alpha \beta }a_{E_{1}}^{\alpha \dag
}a_{E_{2}}^{\beta }\left( b_{j\nu }+b_{j\nu }^{\dag }\right) ,
\label{Helph}
\end{align}%
where $\alpha =L,R$ and $b_{j\nu }$ is the phonon annihilation operator. $%
\left\{ A_{i\mu ,j\nu }\right\} $ is the transformation matrix
that relates Cartesian coordinates to normal coordinates, and
$J_{E_{1},E_{2}}^{i\mu ,\alpha \beta }$ is the electron-phonon
coupling constant which can be directly calculated from the
scattering wave-functions~
\begin{equation}
J_{E_{1},E_{2}}^{i\mu ,\alpha \beta }=\int d\mathbf{r}\int d\mathbf{K}%
_{\Vert }\Psi _{E_{1}}^{\alpha \ast }\left(
\mathbf{r},\mathbf{K}_{\Vert }\right) \partial _{\mu }V^{ps}\left(
\mathbf{r},\mathbf{R}_{i}\right) \Psi _{E_{2}}^{\beta }\left(
\mathbf{r},\mathbf{K}_{\Vert }\right) , \label{coupling}
\end{equation}%
where we have chosen to describe the electron-ion interaction with
pseudopotentials $V^{ps}\left( \mathbf{r},\mathbf{R}_{i}\right) $
for each $i$-th ion~\cite{diventra}.

\begin{figure}
\includegraphics[width=7.5cm]{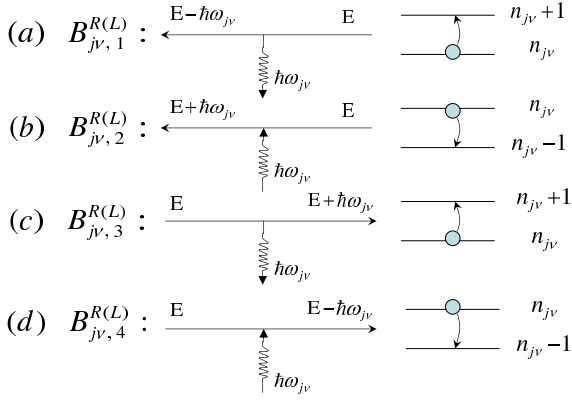}
\caption{Feynman diagrams and corresponding amplitudes (see text)
of the main electron-phonon scattering mechanisms contributing to
the correction of the current and noise.} \label{fig1}
\end{figure}

\begin{figure}
\includegraphics[width=8cm]{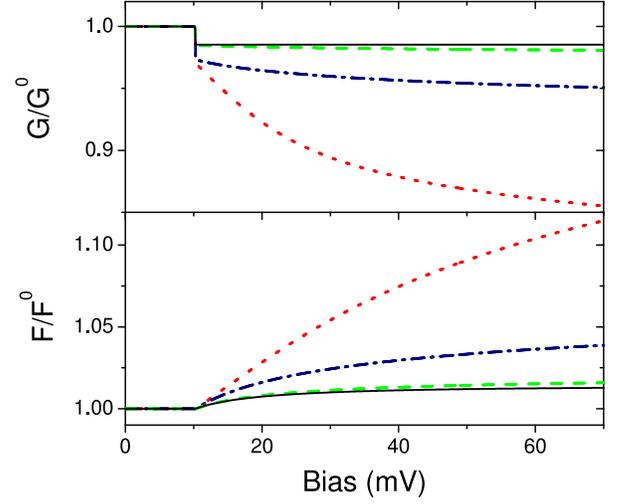}
\caption{Top panel: ratio of the total conductance $G$ of an atomic gold point
contact and its value in the absence of inelastic effects $G^0$ as a function of bias for different values of thermal
current coefficient (see text): $%
A_{th}=10^{-19}$ (dot), $10^{-17}$ (dot-dashed), $10^{-15}$ (dashed), and $%
\infty $ (solid) dyn$/$(sK$^{4}$). Bottom panel: corresponding Fano factor ratio.} \label{fig2}
\end{figure}

We use as unperturbed states of the full system (electron plus
phonon) the
states $|\Psi _{E}^{L(R)};n_{j\nu }\rangle =|\Psi _{E}^{L(R)}\left( \mathbf{%
r,\mathbf{K}_{\Vert }}\right) \rangle \otimes |n_{j\nu }\rangle $, where $%
n_{j\nu }$ is the occupation number of the $j\nu$-th normal mode.
The first-order perturbation to the wave functions is thus
\begin{equation}
|\Phi _{E}^{L(R)};n_{j\nu }\rangle =|\Psi _{E}^{L(R)};n_{j\nu
}\rangle +|\delta \Psi _{E}^{L(R)};n_{j\nu }\rangle ,
\label{per0}
\end{equation}%
where the first-order correction term is
\begin{align}
|\delta \Psi _{E}^{\alpha };n_{j\nu }\rangle & =\lim_{\epsilon
\rightarrow 0^{+}}\sum_{\alpha ^{\prime }=L,R}\sum_{j^{\prime }\nu
^{\prime }}\int
dE^{^{\prime }}D_{E^{^{\prime }}}^{\alpha ^{\prime }}  \notag \\
& \frac{\langle \Psi _{E^{\prime }}^{\alpha ^{\prime
}};n_{j^{\prime }\nu ^{\prime }}|H_{el-vib}|\Psi _{E}^{\alpha
};n_{j\nu }\rangle |\Psi
_{E^{\prime }}^{\alpha ^{\prime }};n_{j^{\prime }\nu ^{\prime }}\rangle }{%
\varepsilon (E,n_{j\nu })-\varepsilon (E^{\prime },n_{j^{\prime
}\nu ^{\prime }})-i\epsilon }\text{,}  \label{perturbation}
\end{align}%
with $D_{E}^{R(L)}$ the partial density of states of left
(right) moving electrons, and $\varepsilon (E,n_{j\nu })=E+\left(
n_{j\nu }+1/2\right) \omega _{j\nu }$ the energy of state
$\left\vert \Psi _{E}^{\alpha };n_{j\nu
}\right\rangle $. Carrying out explicitly the integrals in Eq.~(\ref%
{perturbation}), the nonvanishing corrections to the wave function can
be written as

\begin{eqnarray}
|\delta \Psi _{E}^{\alpha };n_{j\nu }\rangle &=&(B_{j\nu
,1}^{\alpha }+B_{j\nu ,3}^{\alpha })|\Psi _{E+\omega _{j\nu
}}^{\alpha };n_{j\nu
}+1\rangle  \notag \\
&&+(B_{j\nu ,2}^{\alpha }+B_{j\nu ,4}^{\alpha })|\Psi _{E-\omega
_{j\nu }}^{\alpha };n_{j\nu }-1\rangle ,  \label{1storder}
\end{eqnarray}%
where $B_{j\nu ,1}^{\alpha }$, $B_{j\nu ,2}^{\alpha }$, $B_{j\nu
,3}^{\alpha
}$ and $B_{j\nu ,4}^{\alpha }$ correspond to the diagrams depicted in Fig. %
\ref{fig1}. For $\left\vert \delta \Psi _{E}^{R};n_{j\nu
}\right\rangle $, the coefficients are given by:
\begin{eqnarray}
B_{j\nu ,1(2)}^{R} &=&i\pi \sum_{i\mu }\sqrt{\frac{1}{2\omega _{j\nu }}}%
A_{i\mu ,j\nu }J_{E\pm \omega _{j\nu },E}^{i\mu ,LR}D_{E\pm \omega
_{j\nu
}}^{L}  \notag \\
&&\cdot \sqrt{(\delta +\left\langle n_{j\nu }\right\rangle
)f_{E}^{R}(1-f_{E\pm \omega _{j\nu }}^{L})},  \label{bjn1}
\end{eqnarray}%
and

\begin{eqnarray}
B_{j\nu ,3(4)}^{R} &=&-i\pi \sum_{i\mu }\sqrt{\frac{1}{2\omega _{j\nu }}%
}A_{i\mu ,j\nu }J_{E\pm \omega _{j\nu },E}^{i\mu ,RL}D_{E\pm
\omega _{j\nu
}}^{L}  \notag \\
&&\cdot \sqrt{(\delta +\left\langle n_{j\nu }\right\rangle
)f_{E}^{L}(1-f_{E\pm \omega _{j\nu }}^{R})},  \label{bjn2}
\end{eqnarray}%
where $\delta =1$ and "$-$" sign are for the scattering diagrams (a) and (c);
$\delta =0$ and "$+$" sign for diagrams (b) and (d). The average
number of phonons is given by $\left\langle n_{j\nu }\right\rangle
=1/\left[ \exp \left( \omega _{j\nu }/k_{B}T_{w}\right)
-1\right] $ where $T_{w}$ is the local temperature of the
junction~\cite{chen23,chen05}. Similarly, the coefficients in $\left\vert \delta \Psi
_{E}^{L};n_{j\nu }\right\rangle $ have the forms

\begin{equation}
B_{j\nu ,k}^{L}=B_{j\nu ,k}^{R}(L\leftrightharpoons R),
\label{bjn3}
\end{equation}%
where $k=1,\cdots ,4$ ; the notation $\left( L\leftrightharpoons
R\right) $ means interchange of labels R and L.

At $T_{e}=0$ K the first-order correction to the current is thus:

\begin{eqnarray}
I &=&-i\int_{E_{FL}}^{E_{FR}}dE\int d\mathbf{R}\int d\mathbf{K}_{\Vert }%
\tilde{I}_{E,E}^{RR}\cdot   \notag \\
&&\left[ 1-\sum_{j\nu }(\left\vert B_{j\nu ,1}^{R}\right\vert
^{2}+\left\vert B_{j\nu ,2}^{R}\right\vert ^{2})\right] ,
\label{avgI}
\end{eqnarray}%
with $\tilde{I}_{E,E}^{\alpha \beta }\equiv (\Psi _{E}^{\alpha
})^{\ast }\partial _{z}(\Psi _{E}^{\beta })-\partial _{z}(\Psi
_{E}^{\alpha })^{\ast }(\Psi _{E}^{\beta })$. Equation $\left(
\ref{avgI}\right) $ has been simplified by using (i)
$\tilde{I}_{E\pm \omega _{j\nu },E\pm \omega _{j\nu }}^{RR}\simeq
\tilde{I}_{E,E}^{RR}$, valid for energies close to the chemical potentials; and (ii)
$\tilde{I}_{E,E}^{RR}=-\tilde{I}_{E,E}^{LL}$, a direct consequence
of time-reversal symmetry. The current is therefore reduced by
inelastic effects.

Let us now calculate the corresponding correction to shot noise.
We have previously shown that shot noise can be written in terms
of single-particle scattering states as \cite{chen1,johan}

\begin{equation}
S=\int_{E_{FL}}^{E_{FR}}dE\left\vert \int d\mathbf{R}\int d\mathbf{K}\tilde{I%
}_{E,E}^{LR}\right\vert ^{2},  \label{shotnoise}
\end{equation}%
which reduces to the well-known formula $S\varpropto
\sum_{i}T_{i}(1-T_{i})$ when the eigenchannels transmission
probabilities $T_{i}$ are extracted from
the single-particle states with independent transverse momenta~\cite{buttiker,chen1,johan}. Replacing $\left(\ref{per0}\right) $ into $\left(\ref{shotnoise}\right)$
we get
\begin{eqnarray}
S &=&\int_{E_{FL}}^{E_{FR}}dE\left\vert \int d\mathbf{R}\int d\mathbf{K}%
\tilde{I}_{E,E}^{LR}\right\vert ^{2}\cdot \lbrack 1+  \notag \\
&&\sum_{j\nu ;k=1,2}\left( \left\vert B_{j\nu ,k}^{R}\cdot B_{j\nu
,k}^{L\ast }\right\vert ^{2}\right) ]\text{.}  \label{inelshot2}
\end{eqnarray}

Since the summation over vibrational modes contains only positive
terms, shot noise is \emph{enhanced} by electron-phonon inelastic effects in the
quasi-ballistic regime.
Therefore, the Fano factor $F$ normalized to the corresponding value
in the absence of electron-phonon interactions ($F^{0}$) is

\begin{equation}
F/F^{0}=\frac{\int_{E_{FL}}^{E_{FR}}dE\left[ 1+\sum_{j\nu
,k=1,2}\left( \left\vert B_{j\nu ,k}^{R}\cdot B_{j\nu ,k}^{L\ast
}\right\vert ^{2}\right) \right] }{\int_{E_{FL}}^{E_{FR}}dE\left[
1-\sum_{j\nu ,k=1,2}\left\vert B_{j\nu ,k}^{R}\right\vert
^{2}\right] },  \label{FF0}
\end{equation}%
which \emph{increases} with electron-phonon scattering.

Note that due to the orthogonality of phonon states, the absolute
value of the correction to shot noise is smaller than that to the current (cf. Eq.~(%
\ref{avgI}) and Eq.~(\ref{inelshot2})). Note also that
conservation of energy and the Pauli exclusion principle play an
important role. The former dictates an onset bias $V_{c}$ for
inelastic contributions; the latter prohibits the scattering
processes depicted in Fig.~\ref{fig1}(c) and (d) at $T_{e}=0$ K.

These results are illustrated in Fig.~\ref{fig2} where the
inelastic contribution to the conductance and shot noise are
plotted for a gold atom placed in the middle of two bulk gold
electrodes (represented with ideal metals, jellium model,
r$_{\text{s}}\approx $ 3). Details of the calculations of both the
scattering wavefunctions within static density-functional theory
and the vibrational modes for this system can be found in
Refs.~\cite{diventra,chen23}. In the absence of electron-phonon
interactions, the unperturbed differential conductance $G^0$ is
about $1.1$ (in units of $2e^2/h$) and the Fano factor is
$F^{0}\simeq 0.14$~\cite{johan} in the bias range of
Fig.~\ref{fig2}. Inelastic effects cause a discontinuity in the
conductance, and a variation of the Fano factor
ratio~(Eqs.~(\ref{FF0})), at a bias $V_{c}\approx11$ mV,
corresponding to the energy of the lowest longitudinal mode of the
system. In addition, the above inelastic corrections depend on the
local
temperature of the junction $T_{w}$ (see Eqs.~(\ref{bjn1}) through~(\ref%
{bjn3})) which, in turn, is the result of the competition between
the rate of heat generated locally in the nanostructure and the thermal current $I_{th}$
carried away into the bulk
electrodes~\cite{todorov1,todorov2,todorov3,chen23,chen05,yang}. The latter has a temperature
dependence of $I_{th}=A_{th}T^{4}$~\cite{geller}, where the constant $A_{th}$ depends on the details of the
coupling between the local modes of the junction and the modes of the bulk electrodes. At steady state this thermal current has to
balance the power generated in the nanostructure, which is a small fraction of the total
power of the circuit $\frac{V^{2}}{R}$ ($V$ is the bias, $R$ is the resistance)~\cite{todorov1,chen23}.

The larger
$A_{th}$, the larger the heat dissipated into the bulk and, thus, the lower the local temperature
$T_{w}$~\cite{prec1}. In the limit of infinite $A_{th}$, i.e. $T_{w}=0$, at any given bias larger than $V_c$,
electrons can only emit phonons [$\left\langle n_{j\nu }\right\rangle
=0$ in Eqs.~(\ref{bjn1}) and~(\ref{bjn2})]. The inelastic contribution to the conductance and Fano factor, therefore, saturate
to a specific value (see Fig.~\ref{fig2}). We can derive both the bias dependence and this saturation value, to
first order in the bias, as follows.

By equating the thermal current $I_{th}$ to the power generated in the junction, it is easy to show that
$T_{w}=\alpha\sqrt{V}$~\cite{todorov3,length}, where the constant $\alpha $
depends on the details of the thermal contacts between the junction and electrodes. Let us assume for
simplicity a single phonon mode of frequency $\omega$. For $T_{w}> \omega /k_{B}$, we expand
$\left\langle n_{j\nu }\right\rangle \approx k_{B}T_{w}/\omega$ in Eqs.~(\ref{bjn1}) and~(\ref{bjn2}).
From Eq.~(\ref{avgI}) we then get

\begin{equation}
\frac{G}{G^{0}}\simeq 1-\alpha \frac{3}{2}\frac{k_{B}}{
\omega }\gamma _{I}\theta (V-V_{c})\sqrt{V},  \label{GG0}
\end{equation}%
where $\theta (V-V_{c})$ is the Heaviside function; $\gamma
_{I}=\left\vert \left( dI/dV\right) /\left(
dI^{0}/dV\right) \right\vert $ is the relative change in conductance due to inelastic effects at $V_c$ (its value is about 1$\%$ for the specific case,
in agreement with experiments on similar systems~\cite{agrait,chen23}). The inelastic contribution to the
conductance thus decreases with bias as $\sqrt{V}$. This square-root dependence is clear in Fig.~\ref{fig2} for
$A_{th}<10^{-15}$ dyn$/$(sK$^{4})$ which corresponds to temperatures for which the condition $T_{w}> \omega /k_{B}$
is satisfied~\cite{brand}.

The same analysis applied to shot noise leads to
\begin{equation}
\frac{S}{S^{0}}\simeq 1+\alpha ^{2}\left( \frac{k_{B}}{
\omega }\right) ^{2}\gamma _{s}\theta
(V-V_{c})(V-V_c),  \label{SS0}
\end{equation}%
where $\gamma _{S}=\left\vert \left( dS/dV\right) /\left(
dS^{0}/dV\right) \right\vert$ is the relative change of
shot noise due to inelastic effects at $V=V_{c}$ (it is about 0.04 $\%$ for the specific
gold quantum point contact). The inelastic correction to shot noise thus increases linearly with bias for $T_{w}> \omega /k_{B}$.
Consequently, $F/F^{0}\propto \sqrt{V}$ as it is also evident from Fig.~\ref{fig2}.

In the opposite limit of perfect heat dissipation in the bulk electrodes, i.e. for $T_{w}\rightarrow 0$
[see Fig.~\ref{fig2}, $A_{th}\rightarrow \infty $ dyn$/$(sK$^{4}$)],
then from Eqs.~(\ref{bjn1}) and~(\ref{bjn2}) it is easy to prove that $I/I_{0}=1-\theta (V-V_{c})\gamma _{I}(V-V_{c})/V$ and $S/S_{0}= 1+\gamma _{S}\left[(V-V_{c})/V\right]
\theta (V-V_{c})$. Therefore,
\begin{equation}
F/F^{0}=\frac{1+\gamma _{S}\frac{V-V_c}{V}\theta (V-V_{c})}{1-\gamma _{I}\frac{V-V_c}{V}\theta (V-V_{c})},
\end{equation}
which tends to the constant value $F/F^{0}\rightarrow
(1+\gamma _{S})/(1-\gamma _{I})$ as $V >> V_c$.

We have thus shown that the Fano factor depends sensitively on the
efficiency of heat dissipation in nanoscale junctions. It
therefore provides a tool to probe local temperatures and heat
transport mechanisms in these systems. The predictions reported
here should be readily tested experimentally.

We acknowledge partial support from the NSF Grant Nos. DMR-01-33075
and ECS-04-38018. We also thank M. B{\" u}ttiker for useful discussions and
M. Zwolak for help in calculating the reported vibrational modes of the gold
point contact.


\begin{thebibliography}{99}
\bibitem{thermal} At this temperature, and under quasi-ballistic assumptions, thermal noise is
negligible.
\bibitem{prec1} If good thermal transport is assumed between the
junction and the bulk electrodes then the coefficient $A_{th}$ can
be estimated assuming the junction as a weak thermal link with a
given stiffness~\cite{geller}. In this case, $A_{th}\simeq
3.7\cdot 10^{-15}$ dyn$/$(sK$^{4}$)~\cite{chen23}. The
corresponding local temperature is estimated to be $T_w=37$K at
$V=70$ mV~\cite{chen23}. For the other thermal coefficients
reported in Fig.~\ref{fig2}, the local temperature at $V=70$ mV is
$T_w=426$K [$A_{th}= 10^{-19}$ dyn$/$(sK$^{4}$)], $T_w=157$K
[$A_{th}= 10^{-17}$ dyn$/$(sK$^{4})$], and $T_w=51$K
[$A_{th}=10^{-15}$ dyn$/$(sK$^{4})$].
\bibitem{length} The thermal current scales as $l^{-2}$ with wire length $l$ (Ref.~\cite{geller}).
On the other hand, the power dissipated in the junction has weak length dependence for metallic
wires~\cite{todorov3,chen23,yang}; instead it scales as $\exp(-\beta l)$ for insulating junctions, where $\beta$
is a characteristic inverse length of the wire~\cite{chen05}. The local temperature of the wire thus
scales with length as $\sqrt{l}$~\cite{todorov1,smitnanotech} and
$\exp(-\frac{\beta l}{4})$~\cite{chen23,chen05} for metallic and insulating wires,
respectively.
\end{thebibliography}
\end{document}